\documentclass[12pt]{article}

\usepackage{amsmath}
\usepackage{graphicx}
\usepackage{subcaption}
\usepackage{longtable}      
\usepackage{multirow}
\usepackage{hyperref}

\begin{document}

\title{Effective potential of a spinning heavy symmetric top when magnitudes of conserved angular momenta are not equal}
\date{}
\author{V. Tanr{\i}verdi}
\maketitle

\begin{center}
	\vspace{-1cm}
	{tanriverdivedat@googlemail.com}
\end{center}

\begin{abstract}
There are various types of motion of a heavy symmetric top like regular precession, cusp like motion, etc. One of the tools used to understand that motion is effective potential. The effective potential for a spinning heavy symmetric top is studied when magnitudes of conserved angular momenta are not equal to each other.
The dependence of effective potential on conserved angular momenta is analyzed.
This study shows that the minimum of effective potential goes to a constant derived from conserved angular momenta when one of the conserved angular momenta is greater than the other one, 
and it goes to infinity when the other one is greater.
It also shows that the usage of strong or weak top separation does not work adequately in all cases.

\end{abstract}

\section{Introduction}

Motion of a symmetric top can be studied by using either a cubic function or effective potential.
The cubic function is mostly used in works that utilize geometric techniques \cite{Routh, Scarborough, MacMillan, ArnoldMaunder, Groesberg, JoseSaletan},
and effective potential is mostly used in works considering physical parameters \cite{Symon, McCauley, LandauLifshitz, MarionThornton, Taylor}.
In some other works, both the cubic function and effective potential are used \cite{Goldstein, Arnold, Corinaldesi, MatznerShepley, Arya, Greiner, FowlesCassiday}.

Effective potential shows different characteristics when one of the conserved angular momenta greater than the other one or equal to.
One can find different aspects of effective potential in the literature when magnitudes of the conserved angular momenta are equal to each other \cite{Symon, Tanriverdi_abeql}.
However, it is not studied when magnitudes of the conserved angular momenta are not equal to each other except in Greiner's work, 
and his study does not cover different possibilities related to the conserved angular momenta and the minimum of effective potential \cite{Greiner}. 
Studying this topic helps understand the motion of a spinning heavy symmetric top,
and in this study, we will study this case together with the relation between the minimum of effective potential and a constant derived from parameters of gyroscope and conserved angular momenta.

In section \ref{frst}, we will give a quick overview of constants of motion and effective potential.
In section \ref{scnd}, we will study effective potential when magnitudes of the conserved angular momenta are not equal to each other.
Then, we will give a conclusion.
In the appendix, we will compare the cubic function with effective potential.

\section{Constants of motion and effective potential}
\label{frst}

For a spinning heavy symmetric top, Lagrangian is \cite{Goldstein}
\begin{eqnarray}
	L&=&T-U \nonumber \\ 
	&=&\frac{I_x}{2}(\dot \theta ^2 + \dot \phi ^2 \sin^2 \theta)+\frac{I_z}{2}(\dot \psi+\dot \phi \cos \theta)^2-M g l \cos \theta, 
	\label{lagrngn}
\end{eqnarray}
where $M$ is the mass of the symmetric top, $l$ is the distance from the center of mass to the fixed point, $I_x=I_y$ and $I_z$ are moments of inertia, $g$ is the gravitational acceleration, $\theta$ is the angle between the stationary $z'$-axis and the body $z$-axis, $\dot \psi$ is the spin angular velocity, $\dot \phi$ is the precession angular velocity and $\dot \theta$ is the nutation angular velocity. 
The domain of $\theta$ is $[0,\pi]$.
For a spinning symmetric top on the ground $\theta$ should be smaller than $\pi/2$, and if $\theta>\pi/2$, then the spinning top is suspended from the fixed point. 

There are two conserved angular momenta which can be obtained from Lagrangian,
and one can define two constants $a$ and $b$ by using these conserved angular momenta as \cite{Goldstein}
\begin{eqnarray}
	a&=&\frac{I_z}{I_x}(\dot \psi+\dot \phi \cos \theta), \\
	b&=&\dot \phi \sin^2 \theta + a \cos \theta,
\end{eqnarray}
where $a=L_z/I_x$ and $b=L_{z'}/I_x$.
Here, $L_z$ and $L_{z'}$ are conserved angular momenta in the body $z$ direction and stationary $z'$ direction, respectively.

One can define a constant from energy as 
\begin{equation}
	E'=\frac{I_x}{2} \dot \theta ^2 +\frac{I_x}{2} \dot \phi^2 \sin^2 \theta + Mgl \cos \theta \label{eprime},
\end{equation}
and its relation with the energy is $E'=E-I_x^2 a^2/(2 I_z)$.

By using change of variable $u=\cos \theta$, one can obtain the cubic function from \eqref{eprime} as\cite{Goldstein}
\begin{equation}
	f(u)=(\alpha - \beta u)(1-u^2)-(b-a u^2)	
	\label{cubicf}
\end{equation}
which is equal to $\dot u^2$,
where $\alpha=2 E'/I_x$ and $\beta=2 Mgl/I_x$.
This cubic function can be used to find turning angles.

From $E'=I_x \dot \theta ^2/2+U_{eff}$ \cite{LandauLifshitz}, it is possible to define an effective potential
\begin{equation}
	U_{eff}(\theta)= \frac{I_x}{2}\frac{(b-a \cos \theta)^2}{\sin^2 \theta}+Mgl \cos \theta.
	\label{ueff}
\end{equation}

\begin{figure}[!h]
	\begin{center}
		\includegraphics[width=7cm]{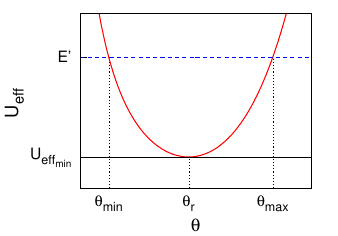}
		\caption{General structure of $U_{eff}(\theta)$ and $E'$. $\theta_{min}$ and $\theta_{max}$ show turning angles, and $\theta_r$ represents the angle where minimum of $U_{eff}$ occurs.
		Curve (red) shows $U_{eff}$, dashed (blue) line shows $E'$ and horizontal continious (black) line shows the minimum of $U_{eff}$.
		}
		\label{fig:ueffg}
	\end{center}
\end{figure}

By using the derivative of $U_{eff}$ with respect to $\theta$
\begin{equation}
	\frac{d U_{eff}(\theta)}{d \theta}= \frac{I_x}{\sin^3 \theta} \left[ (b-a \cos \theta)(a-b \cos \theta)- \frac{Mgl}{I_x} \sin^4 \theta \right],
	\label{dueff}
\end{equation}
it is possible to find the minimum of $U_{eff}$.
The factor $\sin \theta$ is equal to zero when $\theta$ is equal to $0$ or $\pi$, and effective potential goes to infinity at these angles.
The root of equation \eqref{dueff} is between $0$ and $\pi$, and it will be designated by $\theta_r$ giving the minimum of effective potential,
and it can be found numerically.
Then, the form of effective potential is like a well. 
The general structure of $U_{eff}$ together with $E'$ can be seen in figure \ref{fig:ueffg}.

By using equation \eqref{dueff}, one can write \cite{Goldstein}
\begin{equation}
	\dot \phi^2 \cos \theta- \dot \phi a+\frac{Mgl}{I_x}=0.
\end{equation}
The root of this equation can also be used to obtain the minimum of $U_{eff}$.
By using the discriminant of this equation, 
one can define a parameter $\tilde a=\sqrt{4 Mgl /I_x}$ to make a disrimination between "strong top" (or fast top) where $a> \tilde a$ and "weak top" (or slow top) where $a< \tilde a$ \cite{KleinSommerfeld, Tanriverdi_abdffrnt}. 

The position of the minimum and the shape of $U_{eff}$ can be helpful in understanding the motion.
If $E'$ is equal to the minimum of $U_{eff}$ then the regular precession is observed.
If $E'$ is greater than the minimum of $U_{eff}$, like figure \ref{fig:ueffg}, the intersection points of $E'$ and $U_{eff}$ give turning angles.
And, symmetric top nutates between these two angles periodically.
There can be different types of motion, and some of these motions can be determined by using relations between $E'$ \& $Mglb/a$ and $a$ \& $b$ when $|a|\ne|b|$ \cite{Tanriverdi_abdffrnt}.

\section{Effective potential}
\label{scnd}

The relation between $a$ and $b$ can affect effective potential.
There are three possible relation between $a$ and $b$: $|a|>|b|$, $|a|<|b|$ and $|a|=|b|$. 
We will consider two different possibilities, $|a|>|b|$ and $|a|<|b|$, to study effective potential since the third one is studied previously, i.e. $|a|=|b|$ \cite{Symon, Tanriverdi_abeql}.
We will give examples to studied cases, and for examples, the following constants will be used: $Mgl=0.068 \,J$, $I_x=0.000228 \,kg \,m^{2}$ and $I_z=0.0000572 \,kg \,m^{2}$.

\subsection{Effective potential when $|a|>|b|$}

In this section, we will study the case when $|a|>|b|$.
After factoring equation \eqref{dueff}, it can be written as
\begin{equation}
	\frac{d U_{eff}(\theta)}{d \theta}= \frac{a^2 I_x}{ \sin^3 \theta} \left[ (\frac{b}{a}- \cos \theta)(1- \frac{b}{a} \cos \theta)- \frac{Mgl}{I_x a^2} \sin^4 \theta \right].
	\label{dueff3}
\end{equation}
The angle, making the terms in the parentheses zero, gives the minimum of effective potential.
If $|a|>|b|$, the second term in the parentheses is always negative, and then $b/a-\cos \theta$ should be positive for the root.
Therefore, the inclination angle should satisfy $\pi>\theta>\arccos b/a$.
In the limit where $a$ goes to infinity, $\theta_r$ goes to $\arccos b/a$.
In $a$ goes to zero limit, $b$ should also go to zero since $|a|>|b|$, then the first term goes to zero (see equation \eqref{dueff}) and the second term should also go to zero for the root which is possible when $\theta_r$ goes to $\pi$.
If both $a$ and $b$ are negative or positive, $\theta_r$ is between $\pi/2$ and $\pi$ when $|a|$ is close to zero, and it is between $0$ and $\pi/2$ when $|a|$ and $|b|$ are great enough.
If only one of them is negative, then $\theta_r$ is always greater than $\pi/2$.

When $b=0$, in $|a|$ goes to infinity limit $\theta_r$ goes to $\pi/2$, and $a$ goes to zero limit does not change and remains as $\pi$.

These shows that $\theta_r \in (\arccos b/a,\pi)$. 
If $b/a$ goes to $1$, then $\arccos b/a$ goes to $0$.
Therefore, $\theta_r$ can take values between $0$ and $\pi$ depending on signs of $a$ and $b$, the ratio $b/a$ and greatness of $a$ and $b$.

Now, we will consider the change of $U_{eff_{min}}$ when $|a|>|b|$.
We have seen that as $|a|$ goes to zero, $\theta_r$ goes to $\pi$ .
Then, it can be seen from equation \eqref{ueff} that $U_{eff_{min}}$ goes to $-Mgl$ as $|a|$ goes to zero.
As $|a|$ goes to infinity $\theta_r$ goes to $\arccos b/a$, then $U_{eff_{min}}$ goes to $Mgl b/a$ from below.
Then, $Mglb/a$ is always grater than $U_{eff_{min}}$ when $|a|>|b|$.

\begin{figure}[!h]
    \centering
    \begin{subfigure}[b]{0.3\textwidth}
        \centering
        \includegraphics[width=\linewidth]{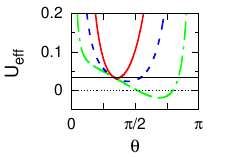}
        \caption{$U_{eff}$}
    \end{subfigure}
    \begin{subfigure}[b]{0.3\textwidth}
        \centering
        \includegraphics[width=\linewidth]{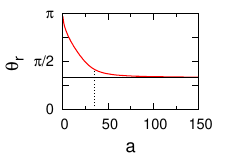}
        \caption{$\theta_r$}
    \end{subfigure}
    \begin{subfigure}[b]{0.3\textwidth}
        \centering
        \includegraphics[width=\linewidth]{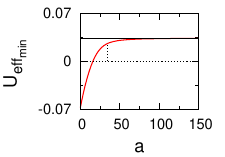}
        \caption{$U_{eff_{min}}$}
    \end{subfigure}

    \caption{$U_{eff}$, change of $\theta_r$ with respect to $a$ and change of $U_{eff_{min}}$ with respect to $a$.
    a) Three different effective potentials: $a=10 \, \mathrm{rad\,s^{-1}}$ (green dashed-dotted), 
    $a=30 \, \mathrm{rad\,s^{-1}}$ (blue dashed), 
    $a=60 \, \mathrm{rad\,s^{-1}}$ (red continuous), all with $b/a=0.5$. 
    Black line shows $Mglb/a$.
    b) Change of $\theta_r$ vs $a$ at constant $b/a=0.5$. Black line shows $\arccos(b/a)=1.05$. Vertical dotted line: $\tilde a$.
    c) Change of $U_{eff_{min}}$ vs $a$ at constant $b/a=0.5$. Black line shows $Mglb/a$. Vertical dotted line: $\tilde a$.}
    \label{fig:ueffg5}
\end{figure}

As an example, we will consider that there is a constant ratio between $a$ and $b$: $b/a=0.5$.
In figure \ref{fig:ueffg5}(a), three different effective potentials for three different $a$ values are shown together with $Mglb/a$.
In this figure, it can be seen that the form and magnitude of the minimum of $U_{eff}$ are changing as $a$ changes, and it can also be seen that $\theta_r$ is also changing.
In figure \ref{fig:ueffg5}(b), it can be seen that $\theta_r$ takes very close values to $\pi$ for very small values of $a$ and goes to $\arccos 0.5=1.05 \,rad$ as $a$ increases.
In figure \ref{fig:ueffg5}(c), it can be seen that the minimum of $U_{eff}$ takes very close values to $-Mgl$ when $a$ is small, and it goes to $Mglb/a$ as $a$ goes to infinity.
These are consistent with previous considerations.

It can be considered that there is a shift in the behaviour of $\theta_r$ and $U_{eff_{min}}$ near $a=\tilde a$.             
But this shift is not sudden, and one can say that the usage $\tilde a$ gives an approximate separation when $|a|>|b|$.

In some cases, $Mgl$ can be negative and there are some differences in effective potential in these cases.
When $Mgl$ is negative, the second term in equation \eqref{dueff3} becomes positive, and then $\arccos b/a> \theta > 0$ for the root.
In the limit where $a$ goes to infinity, again $\theta_r$ goes to $\arccos b/a$.
In $a$ goes to zero limit, $\theta_r$ goes to $0$.
These show that the interval for the minimum of effective potential changed from $(\arccos b/a,\pi)$ to $(0,\arccos b/a)$ when $Mgl$ changed sign from positive to negative.
If both $a$ and $b$ are negative or positive, $\theta_r$ is between $0$ and $\pi/2$.
If only one of them is negative, then $\theta_r$ can be greater than $\pi/2$ when $|a|$ is great enough.
The minimum of $U_{eff}$ goes to $-|Mgl|$ when $a$ goes to $0$, and it goes to $-|Mgl| b/a$ when $a$ goes to infinity when $Mgl$ is negative.

\subsection{Effective potential when $|b|>|a|$}

In this section, we will study the case when $|b|>|a|$.
After factoring equation \eqref{dueff} in another way, it can be written as 
\begin{equation}
	\frac{d U_{eff}(\theta)}{d \theta}= \frac{b^2 I_x}{\sin^3 \theta}  \left[ (1-\frac{a}{b} \cos \theta)(\frac{a}{b} - \cos \theta)- \frac{Mgl}{I_x b^2} \sin^4 \theta \right].
	\label{dueff2}
\end{equation} 
Similar to the previous case, the first term should be positive, and $a/b- \cos \theta$ should be positive when $|b|>|a|$ for the root, and then $\pi>\theta>\arccos a/b$.
In $b$ goes to infinity limit, the second term in the parentheses goes to zero.
Then, as $|b|$ goes to infinity, $\theta_r$ should go to $\arccos a/b$.
In $b$ goes to zero limit, $\theta_r$ goes to $\pi$ which can be seen from equation \eqref{dueff} similar to the previous section. 
Then, $\theta_r$ goes to $\pi$ when $b$ goes to zero, and it goes to $\arccos a/b$ when $|b|$ goes to infinity.

When $a$ and $b$ are both positive or negative, as $|b|$ increases from zero to infinity, $\theta_r$ decreases from $\pi$ to $\arccos a/b<\pi/2$.
If only one of them is positive, then $\theta_r$ is always greater than $\pi/2$ and shows a similar decrease to both positive or negative cases.

When $a=0$, as $|b|$ goes to infinity $\theta_r$ goes to $\pi/2$ and it goes to $\pi$ as $|b|$ goes to $0$.

Similar to the previous case, $\theta_r$ can take values between $0$ and $\pi$ depending on signs of $a$ and $b$, the ratio $a/b$ and greatness of $a$ and $b$.

The magnitude of the minimum of $U_{eff}$ changes with respect to $b$.
In $b$ goes to zero limit, $U_{eff_{min}}$ goes to $-Mgl$ since $\theta_r$ goes to $\pi$.
In $b$ goes to infinity limit, $\theta_r$ goes to $\arccos a/b$, and then the minimum of $U_{eff}$ goes to infinity with $I_x b^2(1-(a/b)^2)/2$.

\begin{figure}[!h]
    \centering
    \begin{subfigure}[b]{0.3\textwidth}
        \centering
        \includegraphics[width=\linewidth]{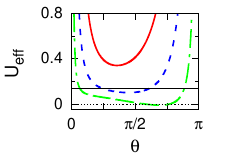}
        \caption{$U_{eff}$}
        \label{fig:ueffg4a}
    \end{subfigure}
    \begin{subfigure}[b]{0.3\textwidth}
        \centering
        \includegraphics[width=\linewidth]{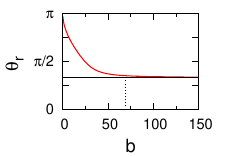}
        \caption{$\theta_r$}
        \label{fig:ueffg4b}
    \end{subfigure}
    \begin{subfigure}[b]{0.3\textwidth}
        \centering
        \includegraphics[width=\linewidth]{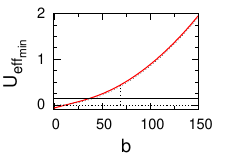}
        \caption{$U_{eff_{min}}$}
        \label{fig:ueffg4c}
    \end{subfigure}

    \caption{$U_{eff}$, change of $\theta_r$ with respect to $b$ and change of $U_{eff_{min}}$ with respect to $b$.
    a) Three different effective potentials: $b=10 \,\mathrm{rad\,s^{-1}}$ (green dashed-dotted), 
    $b=30 \,\mathrm{rad\,s^{-1}}$ (blue dashed), 
    $b=60 \,\mathrm{rad\,s^{-1}}$ (red continuous) with $a/b=0.5$. Black line shows $Mglb/a$. 
    b) Change of $\theta_r$ with respect to $b$ for constant $a/b=0.5$ ratio (red curve). Black line shows $\arccos(a/b)=1.05 \,\mathrm{rad}$. Vertical dotted line: $b=2\tilde a$. 
    c) Change of $U_{eff_{min}}$ with respect to $b$ for constant $a/b=0.5$ ratio (red curve). Black line shows $Mglb/a$. Vertical dotted line: $b=2\tilde a$. Dotted curve shows $I_x b^2(1-(a/b)^2)/2$.}
    \label{fig:ueffg6}
\end{figure}

For examples, similar to the previous case, a constant ratio between $a$ and $b$ is considered: This time $a/b=0.5$.
In figure \ref{fig:ueffg6}(a), three different effective potentials for three different $b$ values are shown similar to the previous section.
In this figure, there are some similarities and differences from figure \ref{fig:ueffg5}(a).
One can see that $\theta_r$ is also different for different $b$ values similar to the previous section.
It can be seen that as $b$ takes different values, the form and magnitude of the minimum of $U_{eff}$ becomes different similar to previous case, and it can be greater than $Mglb/a$, unlike the previous case.
In figure \ref{fig:ueffg6}(b), it can be seen that for very small values of $b$, $\theta_r$ is close to $\pi$ and it goes to $\arccos 0.5=1.05 \,rad$ as $b$ increases.
In figure \ref{fig:ueffg6}(c), it can be seen that the minimum of $U_{eff}$ is close to $-Mgl$ if $b$ is small, and it goes to infinity with $I_x b^2(1-(a/b)^2)/2$ as $b$ goes to infinity.
These are the expected results from the explanations given above.

By considering these results, it can be said that $Mglb/a$ is not important differently from $|a|>|b|$ case.
From figures \ref{fig:ueffg6}(b) and \ref{fig:ueffg6}(c), one can say that the shift in the behaviour of $\theta_r$ and $U_{eff_{min}}$ does not take place around $a=\tilde a$, and the usage of $\tilde a$ for seperation is not suitable when $|b|>|a|$.

When $Mgl$ is negative, the second term in equation \eqref{dueff3} becomes positive, and then in this case, $a/b-\cos \theta$ should be negative which is possible when $\arccos a/b> \theta > 0$.
In the limit where $b$ goes to infinity, again $\theta_r$ goes to $\arccos a/b$.
In $b$ goes to zero limit, $\theta_r$ goes to $0$.
Similar to the previous case, the interval for the minimum of effective potential changed from $(\arccos b/a,\pi)$ to $(0,\arccos b/a)$.
If both $a$ and $b$ are negative or positive, $\theta_r$ is between $0$ and $\pi/2$.
If only one of them is negative, then $\theta_r$ can be greater than $\pi/2$ when $|b|$ goes to infinity, and $\theta_r$ goes to $0$ as $b$ goes to zero.
When $a=0$, in $|b|$ goes to infinity limit $\theta_r$ goes to $\pi/2$, and $|b|$ goes to zero limit does not change and remains as $0$.
If $Mgl$ is negative, the minimum of $U_{eff}$ goes to $-|Mgl|$ when $b$ goes to $0$, and it goes to infinity as $|b|$ goes to infinity.

\section{Conclusion}

Effective potential can be helpful in understanding the motion of a symmetric top in different ways.
$E'$ should be equal to or greater than the minimum of $U_{eff}$ for physical motions.
By using the limits given in section \ref{scnd}, one can say that the regular precession takes place at greater angles when $a$ and $b$ are small, 
and as $a$ and $b$ increase, it takes place at smaller angles.
To observe regular precession smaller than $\pi/2$, $a$ and $b$ should have the same sign and have greater magnitudes.
The limiting angle when $|a|$ or $|b|$ goes to infinity can be found by using inverse cosine of $b/a$ and $a/b$ when $|a|>|b|$ and $|b|>|a|$, respectively.
If $E'$ is greater than the minimum of $U_{eff}$, then different types of motions can be seen \cite{Tanriverdi_abdffrnt}.
These motion will take place closer angles to $\theta_r$ when $E'$ is close to the minimum of $U_{eff}$,
and by considering signs and magnitudes of $a$ and $b$ one can have an opinion on the angles where the motion takes place.

If $a$ and/or $b$ are small, then there can be a high asymmetry in the form of $U_{eff}$.
From the definitions of $U_{eff}$ and $E'$, one can say that $\dot \theta$ is propotional to the difference $E'-U_{eff}(\theta)$ for a specific $\theta$ value.
Therefore, one can say that as $\theta$ increases from $\theta_{min}$ to $\theta_r$, the change in $\dot \theta$ is gradual, and as $\theta$ increases from $\theta_{r}$ to $\theta_{max}$, the change in $\dot \theta$ is more rapid when $a$ and/or $b$ are small.
As $\theta$ changes from $\theta_{max}$ to $\theta_{min}$, this change in $\dot \theta$ is firstly rapid and then gradual.

If $a$ and $b$ are great enough and the difference $E'-U_{eff_{min}}$ is small enough, then the asymmetry in $U_{eff}$ can be ignored.
In these cases, one can make an approximation and find an exact solution for this approximation \cite{Goldstein, Arnold}.
This approximation works better when the asymmetry in $U_{eff}$ is least.

We have seen that comparison of $|a|$ with $\tilde a$ can be used when $|a|>|b|$ for an approximate seperation, and it is not suitable when $|b|>|a|$.
But comparison between $|b|$ and $\tilde a$ can be used when $|b|>|a|$, and if it is used, one should use a naming other than "strong top" or "weak top". 
We should note that comparison of $|a|$ with $\tilde a$ is very useful when $|a|=|b|$  \cite{Tanriverdi_abeql}.

Another thing that should be taken into account is the relation between $Mglb/a$ and $E'$ \cite{Tanriverdi_abdffrnt}.
This study has shown that the minimum of $U_{eff}$ is always smaller than $Mglb/a$ when $|a|>|b|$, which shows that one can always observe all possible motions when $|a|>|b|$.
On the other hand, $Mglb/a$ can be greater than or smaller than the minimum of $U_{eff}$ when $|b|>|a|$.

These results show that effective potential has different advantages over the cubic function in understanding the motion of a spinning heavy symmetric top.
However, the cubic function is still important since it is better for proofs.

\section{Appendix}

There is an alternative to effective potential: the cubic function given in equation \eqref{cubicf}.

Here, we will compare the cubic function with effective potential.
The cubic function is equal to $\dot u^2$, and its roots give the points where $\dot u=0$.
$\dot \theta$ is equal to zero at two of these three points, and the third root is irrelevant to turning angles.
Then, one can use the cubic function to obtain turning angles. 
If these two roots are the same, i.e. double root, then one can also say that this case gives regular precession.
These turning angles can also be obtained from effective potential by using $E'=U_{eff}(\theta)$.
And, if $E'=U_{eff_{min}}$ then the regular precession is observed as explained above.

On the other hand, there is not any correspondence between the minimum of $U_{eff}$ and the maximum of $f(u)$.
The reason for this is the multiplication with $1-u^2$ during the change of variable.
Then, $f(u)$ can not be used to make further analyses similar to $U_{eff}$, given above.

\begin{figure}[!h]
    \centering
    \begin{subfigure}[b]{0.45\textwidth}
        \centering
        \includegraphics[width=\linewidth]{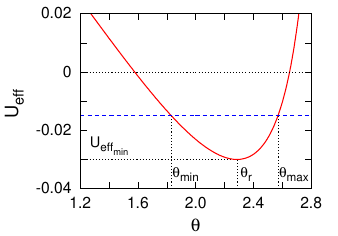}
        \caption{$U_{eff}$}
    \end{subfigure}
    \begin{subfigure}[b]{0.45\textwidth}
        \centering
        \includegraphics[width=\linewidth]{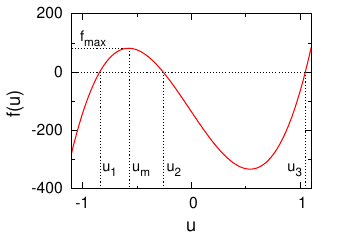}
        \caption{$f(u)$}
    \end{subfigure}

    \caption{$U_{eff}$ and $f(u)$ when $\alpha=575.1 \,\mathrm{s^{-2}}$, 
    $\beta=596.5 \,\mathrm{s^{-2}}$, $a=10 \,\mathrm{rad\,s^{-1}}$ 
    and $b=2 \,\mathrm{rad\,s^{-1}}$.
    a) $U_{eff}$ continuous (red) curve, $E'=-0.0150 \,\mathrm{J}$ dashed (blue) line, 
    $\theta_{min}=1.83 \,\mathrm{rad}$, $\theta_{max}=2.57 \,\mathrm{rad}$, 
    $\theta_r=2.28 \,\mathrm{rad}$ and $U_{eff_{min}}=-0.0299 \,\mathrm{J}$.
    b) $f(u)$ continuous (red) curve, $u_1=-0.841$, $u_2=-0.258$, 
    $u_3=1.05$, $u_m=-0.575$ and $f_{max}=81.6 \,\mathrm{s^{-2}}$.}
    \label{fig:ueff_fu}
\end{figure}

We will consider a case satisfying $\alpha=575.1 \, s^{-2}$, $a=10 \, rad \, s^{-1}$, $b=2 \, rad \, s^{-1}$ as an example.
For the symmetric top with previously given parameters, $\beta$ becomes $ 596.5 \, s^{-2}$.
$U_{eff}$ and $f(u)$ can be seen in figure \ref{fig:ueff_fu}.
One can see that $\theta_{min}=1.83 \, rad$ and $\theta_{max}=2.57 \, rad$ can be obtained from $\arccos(u2)=1.83 \, rad$ and $\arccos(u1)=2.57 \, rad$, respectively.
On the other hand, $\theta_r=2.28 \, rad$ can not be obtained from $\arccos(u_m)=2.18$.

These show that $f(u)$ can be used to obtain turning angles, however, it can not be used to obtain $\theta_r$ where the minimum of $U_{eff}$ occurs.

\end{document}